\documentclass[fleqn,10pt]{wlscirep}
\usepackage[utf8]{inputenc}
\usepackage[T1]{fontenc}
\usepackage{amsmath}
\usepackage{amssymb}
\usepackage{xcolor}
\usepackage{subcaption}
\usepackage[electronic]{ifsym}

\def\dashedrule#1#2#3{{%
\dimen1=#2 \divide\dimen1 by 2
\def\@ruledash{%
\rule{\dimen1}{0pt}%
\rule[0.5ex]{#1}{0.4pt}%
\rule{\dimen1}{0pt}}%
\count1=0
\loop%
\ifnum\count1<#3%
\advance\count1 by 1%
\@ruledash%
\repeat}}

\def\solidline{\dashedrule{2em}{0em}{1}}

\title{%Twin 
Correlated 
transitions in TKE and mass distributions of fission fragments described by 4-D Langevin equation}

\author[1,2]{Mark Dennis Usang}
\author[1,3]{Fedir A. Ivanyuk}
\author[1]{Chikako Ishizuka}
\author[1,4]{Satoshi Chiba}
\affil[1]{Laboratory for Advanced Nuclear Energy, Institute of Innovative Research, Tokyo Institute of Technology, Tokyo 152-8550, Japan.}
\affil[2]{Reactor Technology Center, Technical Support Division, Malaysia Nuclear Agency, Bangi, 43000 Kajang, Malaysia.}
\affil[3]{Nuclear Theory Department, Institute for Nuclear Research, Prospect Nauki 47, 03028 Kiev, Ukraine.}
\affil[4]{Theoretical Division, National Astronomical Observatory of Japan, Mitaka, Tokyo 181-0015, Japan.}

\affil[*]{usang.m.aa@m.titech.ac.jp;  mark\_dennis@nm.gov.my}

\begin{abstract}
We have decomposed to symmetric and asymmetric modes the mass-TKE fission fragment distributions calculated by 4-dimensional Langevin approach and observed how the dominant fission mode and symmetric mode change as functions of $Z^2/\sqrt[3]{A}$ of the fissioning system in the actinides and trans-actinide region.  
As a result, we found that the symmetric mode makes a sudden transition from super-long to super short fission mode around $^{254}$Es.
The dominant fission modes on the other hand, are persistently asymmetric except for $^{258}$Fm, $^{259}$Fm and $^{260}$Md when the dominant fission mode suddenly becomes symmetric although it returns to the asymmetric mode around $^{256}$No.  
These correlated "twin transitions" have been known empirically by Darleane Hoffman and her group back in 1989, but for the first time we have given a clear explanation in terms of a dynamical model of nuclear fission.  
More specifically, since we kept the shape model parameters unchanged over the entire mass region, we conclude that the correlated twin transition emerge naturally from the dynamics in 4-D potential energy surface. 
\end{abstract}
\begin{document}

\flushbottom
\maketitle
\thispagestyle{empty}

%\noindent Please note: Abbreviations should be introduced at the first mention in the main text – no abbreviations lists. Suggested structure of main text (not enforced) is provided below.

\section*{Introduction}

The study of fission by Langevin equation in recent years has had some considerable success \cite{WangYe2018,Eslamizadeh2018,usang2017,Sierk2017,mazurek2017,eslamizadeh2017,usang2016,pahlavani2015,asano2004,karpov2001}, especially in unraveling the physics involved in the fission process. 
%In many calculations of the Langevin equation, we managed to predict the fission observables such as the fission fragment mass yield $Y(A)$ and total kinetic energy TKE. 
%Recent progress include the introduction of 
Recently, we have introduced the 
microscopic mass and friction tensors to improve the calculations of 3-D Langevin equation \cite{usang2016} 
%that has traditionally been using 
instead of the usual
macroscopic mass and friction tensors \cite{wada93,aritomo2013,aritomo2014}. 
%One of the improvements include a better prediction of average TKE, $\left<\mathrm{TKE}\right>$ in comparison to the values from experiments. 
%We also see
With it, we see the average total kinetic energy 
$\left<\mathrm{TKE}\right>$ decreasing with larger excitation energy $E_x$ and the influence of pairing at smaller $E_x$ \cite{usang2017}.
There are some deficiencies with the 3-D Langevin model
%even with the inclusion of microscopic mass and friction tensors. 
%If we study the $\left<\mathrm{TKE}\right>$ across several fission fragments, 3-D Langevin with microscopic cross section however failed to observe the expected transition from $^{256}$Fm to $^{258}$Fm from single peak fission yield to double peak fission yield. It also still fails to produce the correct TKE as a function fragment mass $\mathrm{TKE}(A)$. 
because we were unable to observe the expected transition from double peak fission yield of $^{256}$Fm to the single peak fission yield $^{258}$Fm, and the TKE as a function fragment mass $\mathrm{TKE}(A)$ are rather poor.

These two transitions, in terms of the anomalous changes in the fragment mass yield and TKE, are what we wish to explain. It was first observed experimentally by Hoffman \textit{et al.} \cite{hoffman1980} for $^{258}$Fm and were further corroborated by later experiments \cite{hulet1989,hoffman1990-sf}. Hulet \textit{et al.} \cite{hulet1989} even proposed that these transition occur as the fissioning nucleus  splits into double magic fragments and the high TKE seen for $^{258}$Fm are due to the preference for super-short fission modes \cite{brosa1990}. However, there was no clear explanation why the super-short fission modes are preferred at all instead of the super-long fission modes as was common for all the neighboring actinides.

Thus within the two-center shell model shape parameterization we \cite{lang4D} took into account an additional degree of freedom by allowing the independent deformation of fission fragment tips, and this allowed us to improve $\mathrm{TKE}(A)$ even though we were only able to use it in conjunction with macroscopic transport coefficients.
It seems that the improvements of $\mathrm{TKE}(A)$ is due to the strong relationship between the elongation of the fissioning system at the scission point and the TKE \cite{usang-theory4}. 
In 3-D Langevin approach, the dynamical variables are $(z_0/R_0,\delta,\alpha)$ representing the elongation, fragment tip deformation and mass asymmetry respectively.
%in the present Langevin calculation 
%and it has allowed us to solve the mystery regarding the distribution of $\mathrm{TKE}(A)$ which has been poorly modeled with Langevin. 
We assume in 3-D Langevin approach the shape of the fission fragment tips of the left and right fragments to be the same $(\delta=\delta_1=\delta_2)$. At present we are able to introduce an additional degree of freedom. 
Thus for 4-D Langevin approach, the dynamical coordinates are $(z_0/R_0, \delta_1, \delta_2, \alpha)$ to represent the elongation, right fragment tip shape, left fragment tip shape and mass asymmetry.
Unfortunately, so far we are stuck with macroscopic transport coefficients when we are using 4-D Langevin equations. 

We believe that the more commonly seen transition of fragment mass yield that occurred from $^{256}$Fm to $^{258}$Fm and its recovery at larger compound mass (or charge) are correlated to the anomalous transition of the TKE seen from the same nuclei.
In the present work, we use the 4-D Langevin approach with macroscopic transport coefficients for studying the fragment mass and TKE distributions of various fissioning system from Uranium to Rutherfordium.
%We believe that the transition of 
Our aim is to look for the explanation of the transition from the double peak fragment yield of $^{256}$Fm to single peak fragment of $^{258}$Fm and at the same time, to explain the anomalous increase of $\left<\mathrm{TKE}\right>$ in the said transition.
%In the current work we wish study the bi-modal nature of fission and its relationship with total kinetic energy (TKE) of the fission fragments.

%\section{Calculation Scheme}
%subsection{Langevin Equation}
%Since the publication of our recent work on 4-D Langevin equation for the calculation of fission observables \cite{lang4D}, we have studied several nuclide from our initial study on $^{236}$U up to the isotopes of Rutherfordium.

\section*{Results}
\begin{figure}[h]
\centering
\begin{minipage}{0.49\textwidth}
\centering
\includegraphics[width=\textwidth]{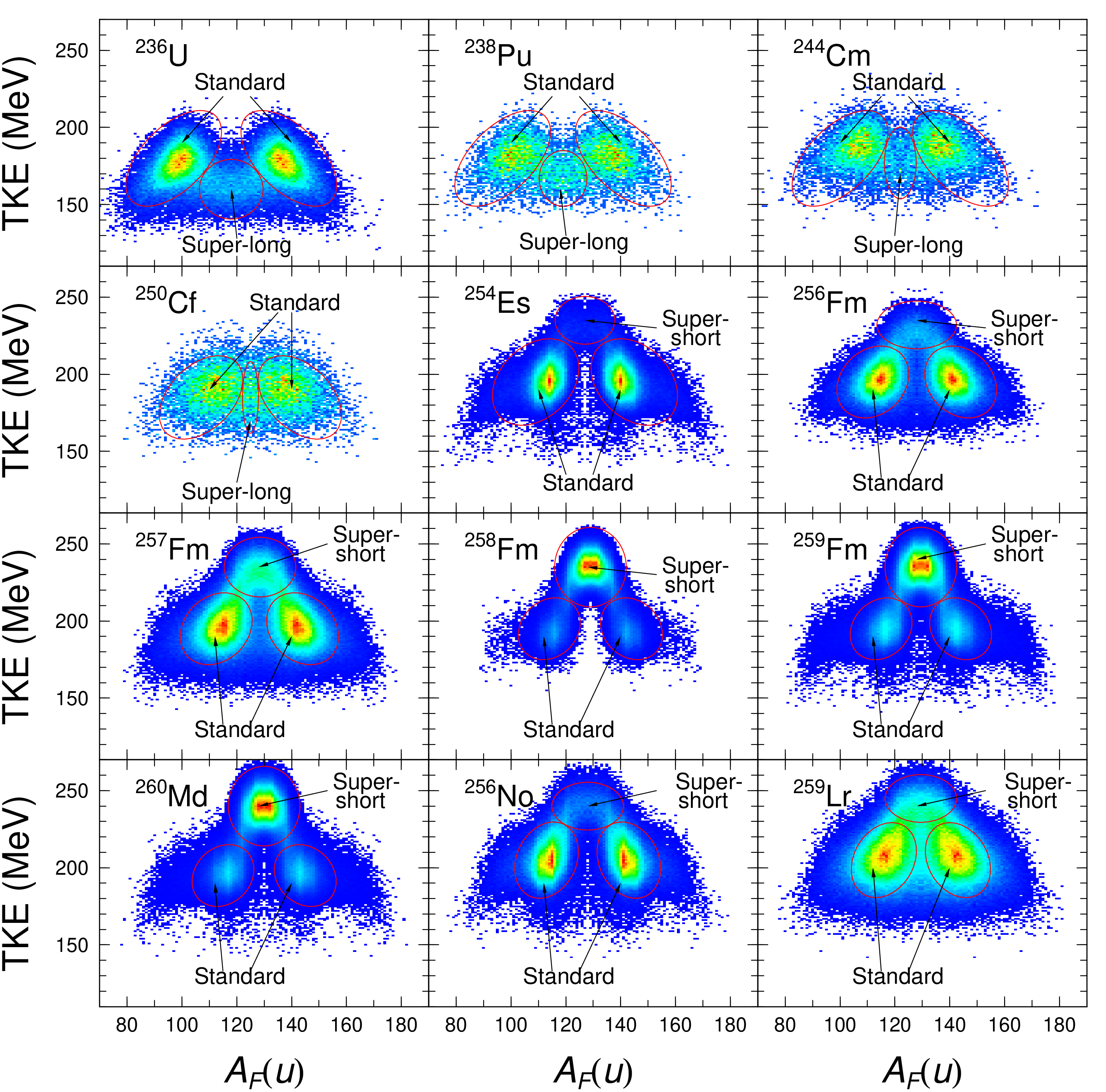}
\caption{Fission fragment TKE profile from $^{236}$U to $^{259}$Lr as a function of fragment mass, $A_F(u)$. The color are linearly scaled from blue to red indicating the counts in arbitrary units.}
\label{u236toLr259}
\end{minipage}
~
\begin{minipage}{0.49\textwidth}
\centering
\includegraphics[width=\textwidth]{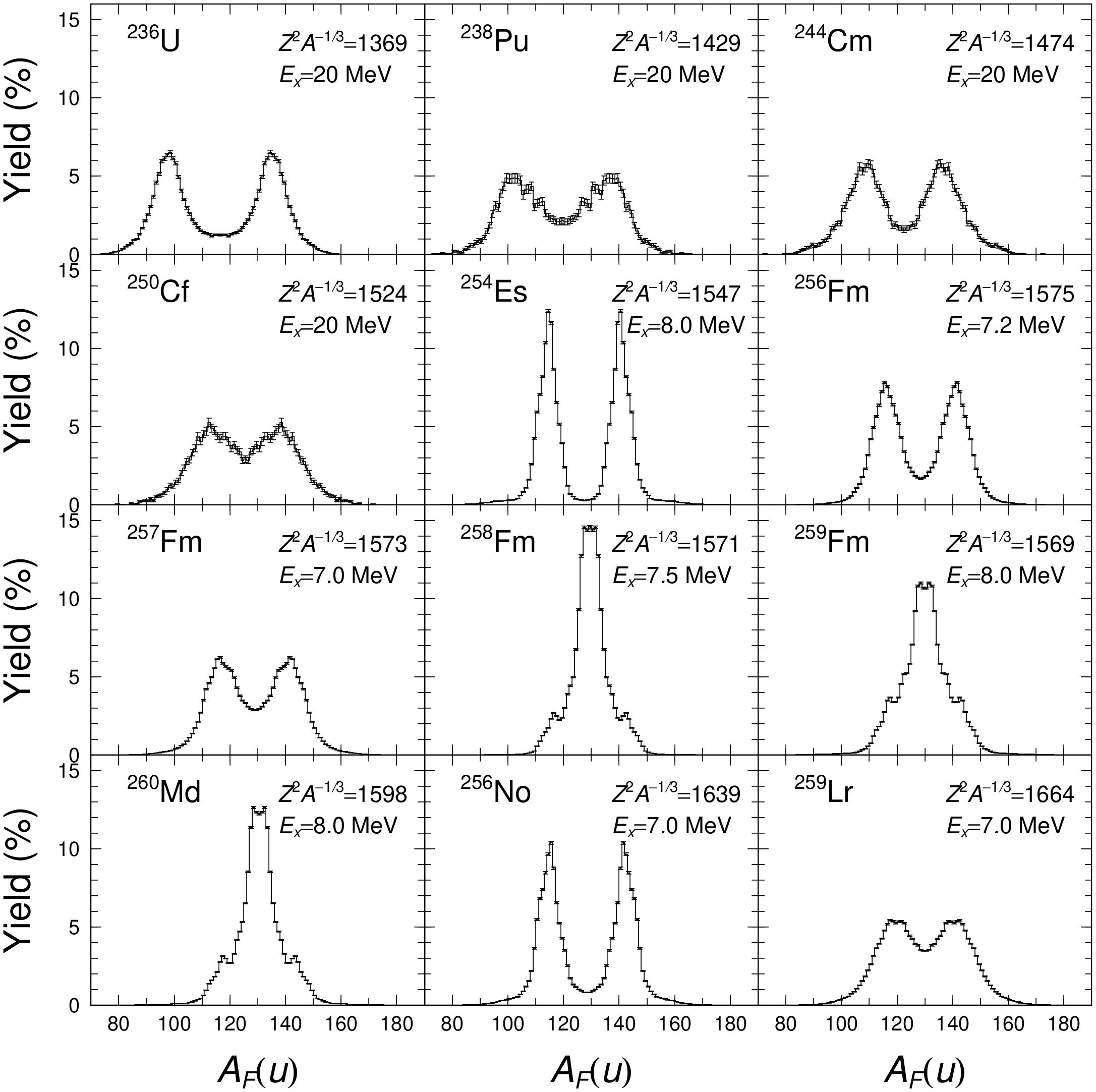}
\caption{ Fission fragment yield from $^{236}$U to $^{259}$Lr as a function of fragment mass $A_F(u)$. The yield are indicated in percentage.
}
\label{yield}
\end{minipage}
\end{figure}

%\textcolor{blue}{It would be good to define here the TKE as the sum of pre-scission kinetic energy and the energy of Coulomb repulsion of future fragments just before scission and mentined that Coulomb energy depends essentially on the distance between fragments centers of mass. Small distance between centers of  mass (short shapes) leads to higher TKE. Large distance between centers of mass (long shapes)leads to smaller TKE. Otherwise it is not clear why the fission event with small TKE are reffered as contributions from super-long mode and the fission events with high TKE are reffered as contributions from super-short mode. The letters in Figs.1,3 should be made of the same size as the main text. I can not read it even with my glases.}
The main observables from Langevin calculation are the fission fragment mass yield and TKE. Fission mass yield is calculated from the statistics of fragment mass given by the value of $\alpha$ at scission. Total Kinetic Energy is calculated from the sum of Coulomb repulsion and pre-scission kinetic energy. 
Brosa \cite{brosa1990} introduced several fission modes associating TKE with the shapes (or length) of the nuclei at scission. 
As the name of these fission modes indicates; the super-short fission modes, standard fission modes and super-long fission modes refers to the length of the nuclide with super-short fission modes being the shortest and super-long fission modes being the longest.
%Up to three levels of \textbf{subheading} are permitted. Subheadings should not be numbered.
In all the nuclei that we managed to calculate with 4-D Langevin, we are able to observe the ever present standard fission modes \cite{brosa1990} manifesting itself in the asymmetric TKE components.
In the lighter fissioning system such as $^{236}$U, we could easily identify the presence of super-long fission modes (smaller TKE) in the symmetric components of the mass distributions. 
On the other hand, the heavier fissioning system such as $^{258}$Fm exhibits super-short fission modes (larger TKE) in the symmetric components.
Snapshots of the TKE profile for a chosen nucleus representing the fissioning system from $^{236}$U all the way up to $^{259}$Lr can be observed in Fig.~\ref{u236toLr259}. 

% \begin{figure}[h]
% \centering
% \begin{subfigure}{0.49\textwidth}
% \centering
% \includegraphics[width=\textwidth]{tke_U236toLr259.eps}
% \caption{\label{u236toLr259} Fission fragment TKE profile from $^{236}$U to $^{259}$Lr as a function of fragment mass, $A_F(u)$. The color are linearly scaled from blue to red indicating the counts in arbitrary units.}
% \end{subfigure}
% ~
% \begin{subfigure}{0.49\textwidth}
% \centering
% %\includegraphics[width=0.45\textwidth]{yield_some2.eps}
% \includegraphics[width=\textwidth]{yield_all.eps}
% \caption{\label{yield} Fission fragment yield from $^{236}$U to $^{259}$Lr as a function of fragment mass $A_F(u)$. The yield are indicated in percentage.
% }
% \end{subfigure}
% \caption{\label{u6toL9} Fission fragment TKE profile and yield from several compound nucleus with increasing charge from $^{236}$U in the top left panel to $^{259}$Lr in the bottom right panel. }
% \end{figure}

%The average value of TKE is roughly proportional to $Z_1Z_2/R$ and TKE$_\mathrm{asy}$ also seems to increase with larger $Z$.
The average value of TKE is roughly proportional to $Z_1Z_2/R$ and it also seems to increase with larger $Z$.
The super-long fission mode slowly diminishes on the symmetric component in tandem with the increasing compound charge until at some point ($^{254}$Es) it suddenly switches to super-short in the symmetric component. 
By drawing an ellipse over the fission modes that we have identified on each TKE profile and took the average TKE in the area spanned by the ellipse for the identified standard fission mode in the asymmetric component, we get TKE$_\mathrm{asy}$.
We see TKE$\mathrm{asy}$ increasing in tandem with the increase of $Z$. 
In a similar fashion we identify TKE$_\mathrm{sym}$ from the super-long or super-short fission modes. From Curium to Californium, it was not clear if any symmetric fission modes are present at all. For such cases, we simply took a narrow band around the symmetric mass as TKE$_\mathrm{sym}$ for the said nuclei.
The symmetric component seems to have exclusively super-short fission modes for all heavier actinides onwards. 
The snapshot of the TKE profile from $^{257}$Fm to $^{259}$Lr in Fig.~\ref{u236toLr259} illustrate these phenomena pretty well.

% \begin{figure}[h]
% \centering
% \includegraphics[width=0.5\textwidth]{Fm257toLr259.eps}
% \caption{\label{fm257onwards} TKE profile for $^{257}$Fm, $^{258}$Fm, $^{259}$Fm, $^{260}$Md, $^{256}$No and $^{259}$Lr TKE profiles respectively beginning from the bottom panel and arranged from left to right.}
% \end{figure}

% \begin{figure}[t]
% \centering
% %\includegraphics[width=0.45\textwidth]{yield_some2.eps}
% \includegraphics[width=0.45\textwidth]{yield_all.eps}
% \caption{\label{yield} Fission fragment yield from $^{236}$U to $^{259}$Lr.}
% \end{figure}

In the case of $^{258}$Fm, $^{259}$Fm and $^{260}$Md, the fission fragments tend to have double magic configuration when they split symmetrically. Due to the preference for symmetric split the only symmetric fission modes available is super-short fission mode, hence it dominates the TKE profile. 
As a consequence, $\left< \mathrm{TKE} \right>$ are also pulled higher. 
%This is the origin of the anomalous TKE often seen in plot such as in Fig. \ref{systematic}. 
Away from the double magic splits, we see that although super short fission modes are still the preferred symmetric fission mode, the asymmetric fission modes dominate instead. 
In the perspective of fission fragment mass yield, this meant that the usual two-peak fragment yield became single-peak for $^{258}$Fm, $^{259}$Fm and $^{260}$Md, and then switched back to double-peak fragment yield.
In Fig.~\ref{yield}, we demonstrate the transition from the double peak $^{256}$Fm to the single peak $^{258}$ and the recovery of double peak fission fragment yield in $^{259}$Lr.

The presence of super-short fission modes for $^{254}$Es are independent from its excitation energy at 8 MeV. For example, at excitation energy of 20 MeV, $^{254}$Es TKE$_\mathrm{asy}$ component is around 195.74 MeV but its $\left< \mathrm{TKE} \right>$ 197.19 MeV. The average TKE is pulled higher by the symmetric TKE components belonging to the super short fission modes. The super-short fission modes component averages around 225.82 MeV. Of course, the majority of the fission events split in an asymmetric manner, but the portion of events that does split symmetrically are due to super-short fission modes. The average TKE of the same nuclide for most fissioning systems, seem to vary with increasing excitation energy. It must be noted however that such variation are usually less than 5 MeV.

\subsection*{Systematics}

\begin{figure}[h]
\begin{minipage}[t]{0.49\textwidth}
\centering
\includegraphics[width=\textwidth]{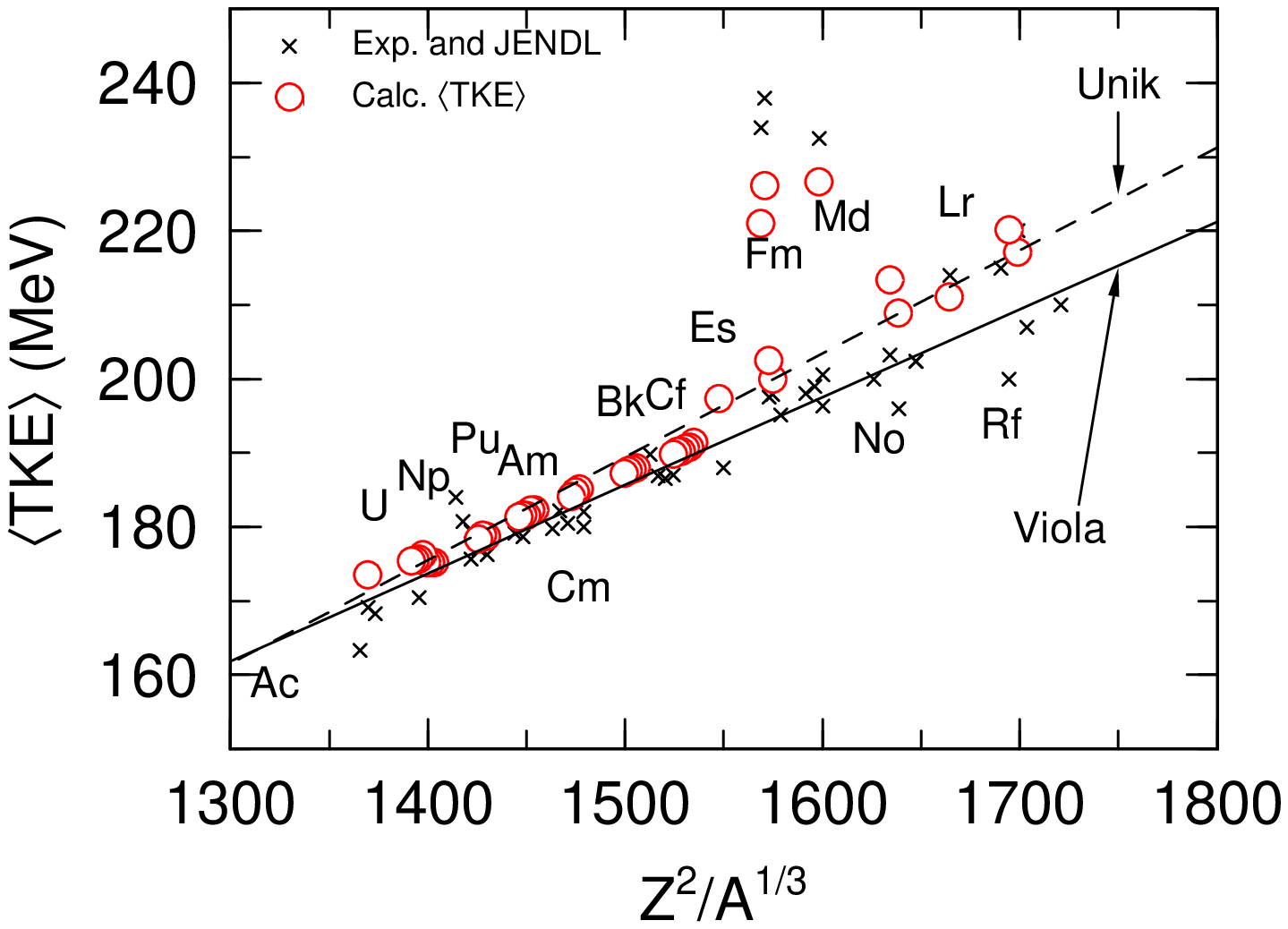}
\caption{\label{systematic} Experimental TKE \cite{hoffman1995,hoffman1996} and evaluated data \cite{JENDL} denoted by ($\times$) compared to calculated $\left<\mathrm{TKE}\right>$ (\textcolor{red}{$\circ$}) as a function of fissioning system.
}
\end{minipage}
\begin{minipage}[t]{0.49\textwidth}
\centering
\includegraphics[width=\textwidth]{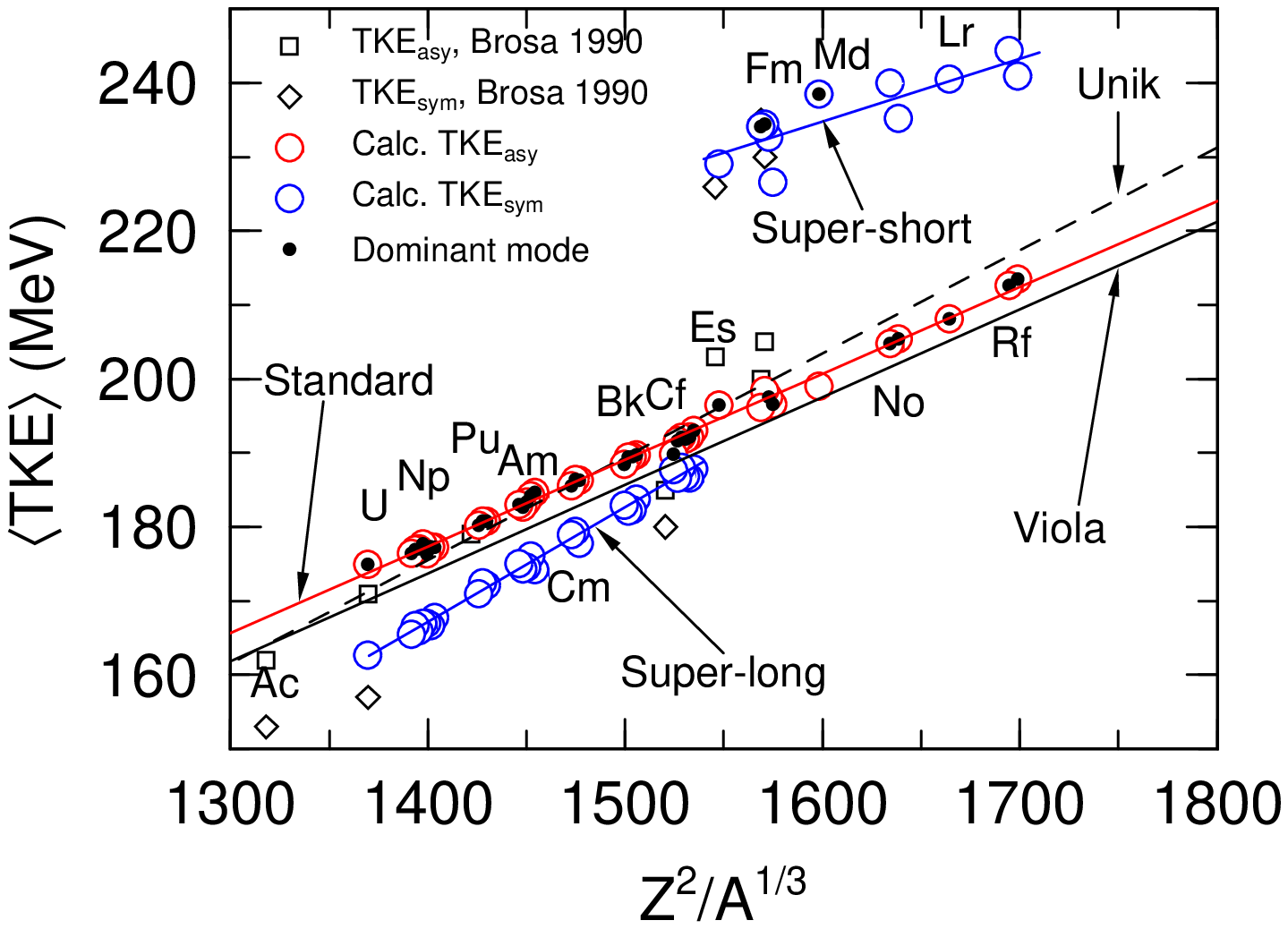}
\caption{\label{systematic-mode} 
Calculated TKE$_{\mathrm{sym}}$ (\textcolor{blue}{$\circ$}) and TKE$_{\mathrm{asy}}$ (\textcolor{red}{$\circ$}) and the associated trend given respectively by (\textcolor{blue}{\solidline}) and (\textcolor{red}{\solidline}).
The dominant fission mode are marked by (\textcolor{black}{$\bullet$}).
Data is plotted as a function of fissioning system.
%These trends are compared with the TKE systematics of \citet{viola1985}(\solidline) and by the systematics of \citet{unik1971} (\sparsedashes). 
}
\end{minipage}
%\caption{\label{syscom} The calculated TKE and its components as a function of fissioning system.}

\end{figure}

We can see from Fig.~\ref{systematic} that while all the other nuclei $\left<\mathrm{TKE}\right>$ seems to follow closely either the Viola systematics \cite{viola1985}% given as,
% \begin{equation}
% \mathrm{TKE}_\mathrm{Viola}=0.1189\left(Z^2/\sqrt[3]{A}\right)+7.3,
% \end{equation}
or the systematics of Unik (Double Energy Experiments) \cite{unik1971},
% \begin{equation}
% \mathrm{TKE}_\mathrm{Unik}=0.1396\left(Z^2/\sqrt[3]{A}\right)-19.9,
% \end{equation}
the three nuclei $^{258}$Fm, $^{259}$Fm and $^{260}$Md clearly deviates away from them.
Both systematics are obtained by taking the linear fit of the TKE as a function of Coulomb repulsion from various fission experiments.
If we take in the trends of the asymmetric TKE components (standard fission modes) from TKE$_\mathrm{asy}$, we get
%\begin{equation}
$\mathrm{TKE}_{STD}=0.1168Z^2A^{-1/3}+13.9$ MeV.
%\end{equation}
It is most interesting to note how close the slope coefficient of Viola value is with the slope coefficient of $\mathrm{TKE}_{STD}$ across the various fissioning system.
The results used in plotting both Fig. \ref{systematic} and Fig. \ref{systematic-mode} are tabulated in the supplementary information and it includes other notes on the particulars of the calculation.

With regards to the symmetric component, the super-long fission mode %$\left<\mathrm{TKE}\right>$ 
approaches that of the standard fission mode in Fig.~\ref{systematic-mode} consistent to what we see in Fig.~\ref{u236toLr259} with the disappearing super-long fission modes. Thus the TKE$_{\mathrm{sym}}$ for the range the fissioning system $1300 <Z^2A^{-1/3} <1550$ tracking super-long fission trends,
%\begin{equation}
%\label{sl}
$\mathrm{TKE}_{SL}=0.1542Z^2A^{-1/3}-48.7$ MeV.
%\end{equation}

% \begin{figure}[h]
% \centering
% \includegraphics[width=0.45\textwidth]{systematic_sym-asy2_20MeV.eps}
% \caption{\label{systematic-mode} 
% Calculated TKE$_{\mathrm{sym}}$ (\textcolor{blue}{$\circ$}) and TKE$_{\mathrm{asy}}$ (\textcolor{red}{$\circ$}) and the associated trend given respectively by (\textcolor{blue}{\solidline}) and (\textcolor{red}{\solidline}).
% The dominant fission mode are marked by (\textcolor{black}{$\bullet$}).
% %These trends are compared with the TKE systematics of \citet{viola1985}(\solidline) and by the systematics of \citet{unik1971} (\sparsedashes). 
% }
% \end{figure}

%%% Consideration for deletion %%%
We could very well see that the slope of $\mathrm{TKE}_{SL}$ predicts the TKE$_{\mathrm{sym}}$ of smaller fissioning system such as $^{227}$Ac but not for indefinitely smaller fissioning system. TKE$_\mathrm{sym}$ given by Brosa \cite{brosa1990} for $^{213}$At and  $^{227}$Ac are 146 MeV and 153 MeV respectively, and TKE$_\mathrm{sym}$ predicted by TKE$_{SL}$ are  137.9 MeV and 151.5 MeV for each fissioning system. Thus for fissioning system that are decreasingly smaller, the steep slope of $\mathrm{TKE}_{SL}$ might taper slightly.
%%%%%%%%%%%%%%%%%%%%%%%%%%%%%%%%%%%%
Systematics from calculated TKE$_{\mathrm{sym}}$ for fissioning system $Z^2A^{-1/3}> 1550$ effectively gives the trend for super-short fission modes,
%\begin{equation}
$\mathrm{TKE}_{SS}=0.0849Z^2A^{-1/3}+99.0$ MeV
%\end{equation}
shows that the super-short TKE is much flatter. 
%We could see from the differences of $\left<\mathrm{TKE}\right>$ from experiments against our calculated $\left<\mathrm{TKE}\right>$ at least for $^{258}$Fm, $^{259}$Fm and $^{260}$Md could be attributed to the deficiency in the current implementation of 4-D Langevin due to the use macroscopic transport coefficients. 
The prediction of the TKE$_{\mathrm{asy}}$ and TKE$_{\mathrm{sym}}$ are quite excellent but there are too much asymmetric fragments in the calculation. This could probably be fixed by adopting the more realistic microscopic transport coefficients. 

\subsection*{The Trajectories}
In order to explain why our calculations are able to reproduce the correlations between mass- and TKE-distributions, the immediate idea would be to analyze the potential energy $U(q)$ for 3D and 4D calculations. This turns out to be very complicated due to the large dimensions involved. Neither does minimizing $U(q)$ with respect to $\delta$ in every $(z_0/R_0,\alpha)$ coordinates gives any useful information because it cannot discriminate between forbidden and allowed fission paths, especially for the heavier actinides. 
It makes some sense to look at how $\delta$ is distributed at scission because the failure of minimization in $\delta$ indicates that the fission paths in $\delta$-space might be crucial for the shape of the fission yield.
%Another way to confirm the dominant fission modes is to look at the distribution of $\delta$.

\begin{figure}[h]
\centering
\begin{minipage}[t]{0.49\textwidth}
\centering
\includegraphics[width=\textwidth]{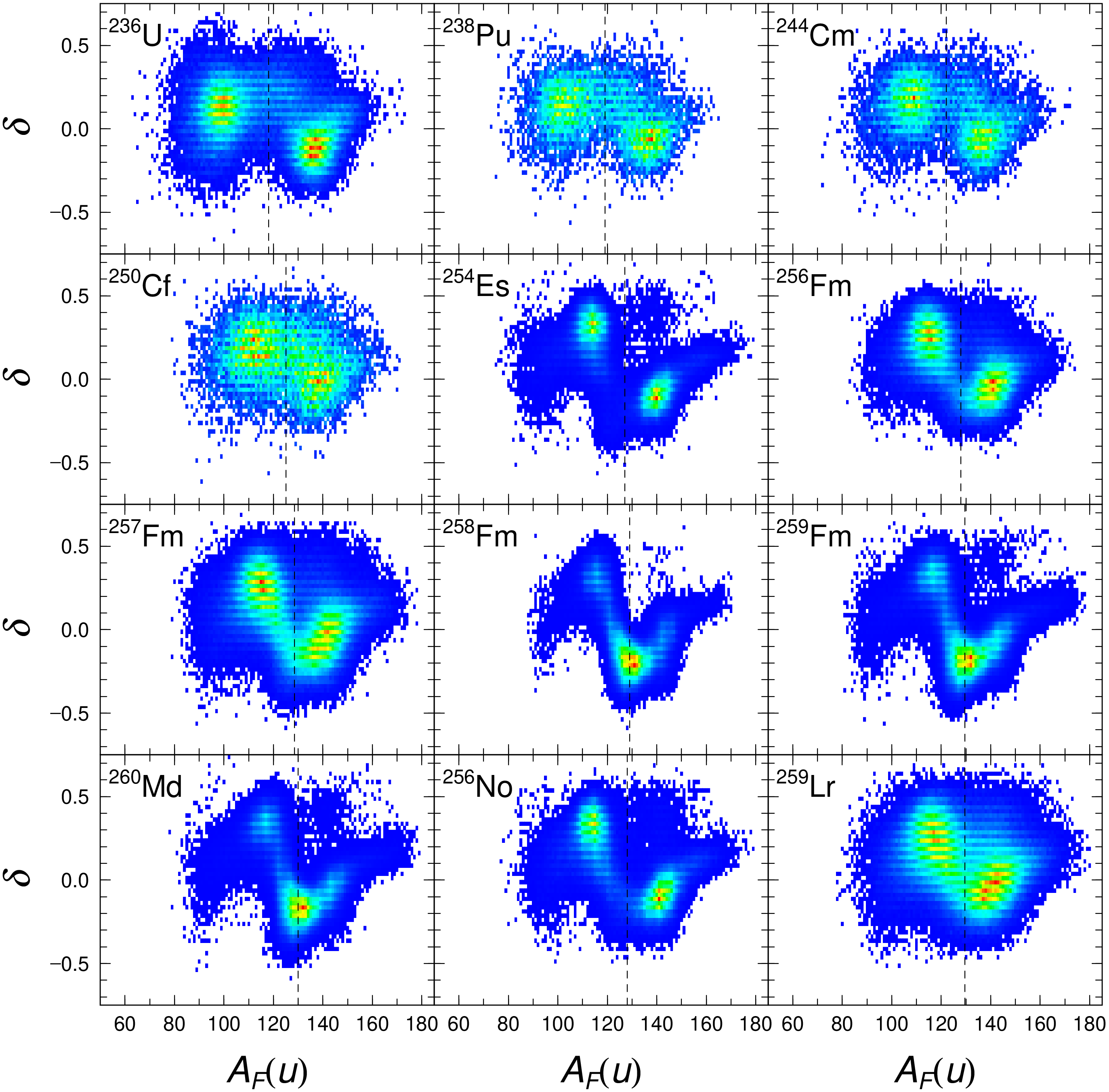}
%\caption{\label{delta-all} Distributions of $\delta$ for $^{256}$Fm to $^{259}$Lr.}
\caption{\label{delta-all} Distributions of $\delta$ as a function of fragment mass, $A_F(u)$. Dashed line to indicate the symmetric fission fragment mass. The colors of the distribution scales linearly from blue to red indicating the increasing counts in arbitrary units.}
\end{minipage}
~
\begin{minipage}[t]{0.49\textwidth}
\includegraphics[width=\textwidth]{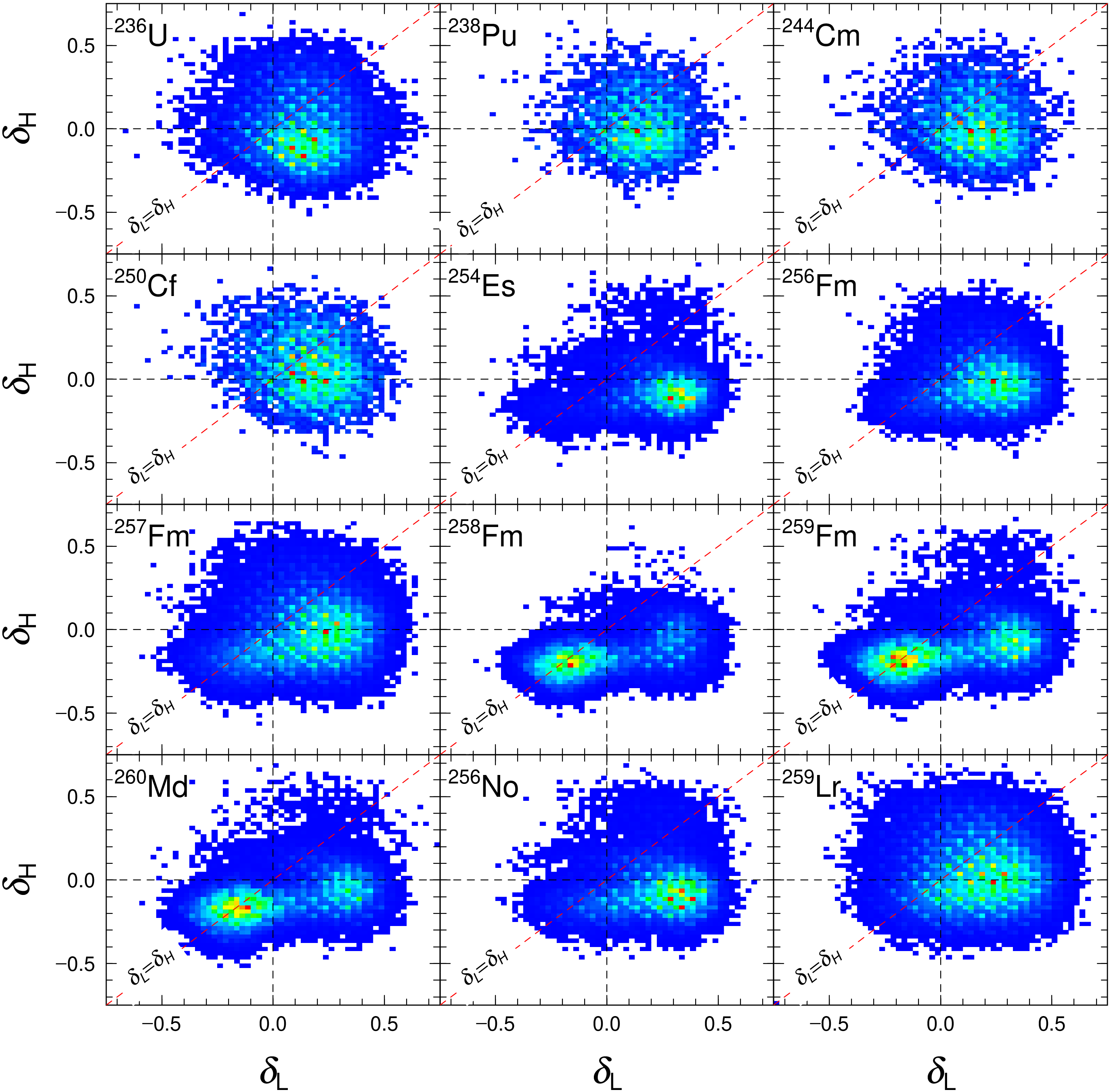}
\caption{\label{fedex} Combinations of $\delta_L$ and $\delta_H$ at scission for selected nuclei from $^{236}$U to $^{259}$Lr , summed over all possible asymmetries of fission fragments. Black dashed line indicate $\delta_L=0$ in x-axis and $\delta_H=0$ in y-axis. Red dashed line indicate $\delta_L=\delta_H$. The colors of the distribution scales linearly from blue to red indicating the increasing counts in arbitrary units.}
\end{minipage}
%\caption{\label{pos} Behaviors of $\delta$ at scission for $^{236}$U to $^{259}$Lr. The color are linearly scaled from blue to red indicating the counts in arbitrary units. }
\end{figure}

Thus we first look at the distribution of $\delta$ with respect to the fission fragment mass.
Positive $\delta$ means that the fragment tip is prolate, negative $\delta$ means that it is oblate  and $\delta=0$ imply that fragments tips are  spherical.  
Comparing Figs. \ref{u236toLr259} and \ref{delta-all} we can see that the standard fission modes correspond to positive $\delta$ for $A_L$ and negative $\delta$ for $A_H$. 
Super long fission modes have positive $\delta$ for both fragments. 
Super short fission modes have negative $\delta$ for both fragments. 
In Fig.~\ref{delta-all}, we see the dominant super short fission modes in the expected single-peak yield nuclei and all double peak yield nuclei has dominant standard fission modes.

%\textcolor{blue}{In Fig.2 'A' denotes the mass number of the whole nuclei, what is quite common. In Figs.1,3,4 it woould be better to use 'Fragment mass number' instead of 'A'.}

With that established, we can probably deduce what happened in the $\delta$-space of $U(q)$ by plotting the combination of $\delta$ for the light and heavy fission fragment denoted each by $\delta_L$ and $\delta_H$. 
Fig.~\ref{fedex} immediately showed us the differences in the $(\delta_L,\delta_H)$ combination for fissioning system smaller than $^{254}$Es against the ones equal to and larger than $^{254}$Es. 
The somewhat symmetrically distributed $(\delta_L,\delta_H)$ combinations of $^{236}$U to $^{250}$Cf meant that in $\delta$-space there is only a single fission path or at least the fission path are very close to each other. 
This fission path seem to lead to standard fission modes. The super-long fission mode fission path might be present but it seems to be very close to the fission path leading to standard fission modes.
It also explain the success of 3-D Langevin model in describing them; after all a single fission path in $\delta$ meant that it was simply unnecessary to go to higher dimension.
However, from $^{254}$Es $(\delta_L,\delta_H)$ combinations became asymmetric. 
In $^{258}$Fm, $^{259}$Fm, $^{260}$Md it is revealed that the asymmetry of $(\delta_L,\delta_H)$ are due to the presence of two fission paths in $\delta$-space. 
The first fission path leads to the usual standard fission modes. The second one leads to the super-short fission modes. 
Due to the multiple fission path and asymmetry of $(\delta_L,\delta_H)$ combinations, 3-D Langevin equations are unable to solve the transition between double-peak fission fragment yield to single peak fission yield. 
Now, with our 4-D Langevin, this is solved.

\begin{figure}[h]
\centering
\includegraphics[width=0.9\textwidth]{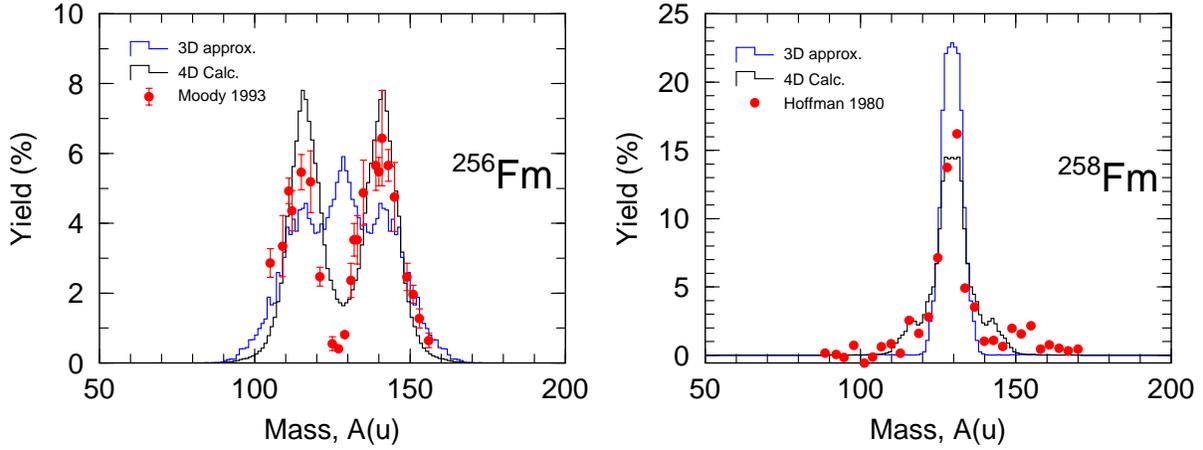}
\caption{4-D Langevin calculated fission yield (\FallingEdge) and the approximation to 3-D Langevin (\textcolor{blue}{\FallingEdge}) for $^{256}$Fm on the left and $^{258}$Fm on the right. Comparison with experimental data (\textcolor{red}{\textbullet}) for $^{256}$Fm \cite{moody} and $^{258}$Fm \cite{hoffman1980} are made for the respective figure from left to right.}
\label{3D-approx}
\end{figure}

We can examine these conjectures regarding the relationship of the fission modes with the combinations of $\delta_L$ and $\delta_H$ by extracting the fission fragment yield for the specific case of $\delta_L \approx \delta_H$. 
All fission events satisfying the condition $ \left| \delta_1 - \delta_2 \right| \leqslant 0.025$ are collected and then we proceed to calculate the fission fragment yield associated with it.
We may consider such procedure an approximation of 3-D Langevin calculations using 4-D Langevin equation. 
The approximated 3-D Langevin fission yield may then be compared with the fission yield calculated with 4-D Langevin fission yield that we had gave in Fig. \ref{yield}. 

In Fig. \ref{3D-approx}, we compare the approximated 3-D Langevin fission yield with the 4-D Langevin fission fragment yield for $^{256,258}$Fm.
Our approximated 3-D Langevin fission yield displayed a three peak structure for $^{256}$Fm.
The symmetric peak should originate from the super-short fission modes meanwhile the asymmetric peak should originate from the standard fission modes.
If we reflect again to the combinations of $\delta_L$ and $\delta_H$ that we see in Fig. \ref{fedex}, we can see most the events associated with the standard fission modes for $^{256}$Fm is far off from the line for $\delta_L=\delta_H$.
Hence the dominance of the symmetric yield in approximated 3-D Langevin fission yield.
In 4-D Langevin calculation of $^{256}$Fm, the super-short fission events are overwhelmed by the statistics from the standard fission events. 
Thus the fission yield by 4-D Langevin gave the familiar asymmetric double peak fission yield as was indicated from the experimental fission yield from Moody \textit{et al.} \cite{moody}.
The experimental fission yield gave an excellent fit of the heavy fragment yield but the valley of the experimental fission yield is much lower. 
The lighter part of the experimental fission yield are also slightly shifted to the lighter mass and the peak height lower.
The slight misalignment of the calculated 4-D Langevin fission yield and experimental fission yield are most likely due to the prompt neutron emission in the experiment.

In the case of $^{258}$Fm, the approximated 3-D Langevin give the correct yield consistent with 4-D Langevin fission yield and experimental fission yield \cite{hoffman1980}. 
One deficiencies that we can see from the approximated 3-D Langevin is the thinness of the yield due to the lack of events with standard fission modes when $\delta_L=\delta_H$.
Apart from the slight misalignment of the peak for the calculated 4-D Langevin fission yield and experimental fission yield by Hoffman \textit{et al.} \cite{hoffman1980} most likely due to neutron emission, we can say that the experimental fission yield is well reproduced.
Thus we now know the origins of the poor transition from $^{256}$Fm to $^{258}$Fm by 3-D Langevin and also perhaps of other methods relying on 3-D potential energy surface.

% \begin{figure}[h]
% \centering
% \begin{subfigure}{0.8\textwidth}
% \centering
% \includegraphics[width=\textwidth]{dhl_fm256-v2.eps}
% \caption{4-D Langevin calculated (|\FallingEdge) fission yield for $^{256}$Fm and the approximation to 3-D Langevin (\textcolor{blue}{|\FallingEdge}). Comparison with experimental data \cite{moody} (\textcolor{red}{\textbullet}).}
% \end{subfigure}
% ~
% \begin{subfigure}{0.8\textwidth}
% \centering
% \includegraphics[width=\textwidth]{dhl_fm258-v2.eps}
% \caption{4-D Langevin calculated (|\FallingEdge) fission yield for $^{258}$Fm and the approximation to 3-D Langevin (\textcolor{blue}{|\FallingEdge}). Comparison with experimental data \cite{hoffman1980} (\textcolor{red}{\textbullet}).}
% \end{subfigure}
% \caption{\textit{Left}: 4-D Langevin distribution of fission observables as function $\delta_L$ in x-axis and $\delta_H$ in y-axis with the yield increasing arbitrarily from blue to red. \textit{Right}: 4-D Langevin calculation fission and the approximation to 3-D Langevin by plotting only 4-D Langevin fission observables with $\delta_L \approx \delta_H$.}
% \end{figure}

% \begin{figure}
% \includegraphics[width=0.5\textwidth]{dhl.eps}
% \caption{\label{fedex} Combinations of $\delta_L$ and $\delta_H$ for selected nuclei from $^{236}$U to $^{259}$Lr. Dashed line indicate $\delta_L=0$ in x-axis and $\delta_H=0$ in y-axis. The color are linearly scaled from blue to red indicating the counts in arbitrary units.}
% \end{figure}

\section*{The Model}

\begin{figure}[h]
\centering
\includegraphics[width=0.8\textwidth]{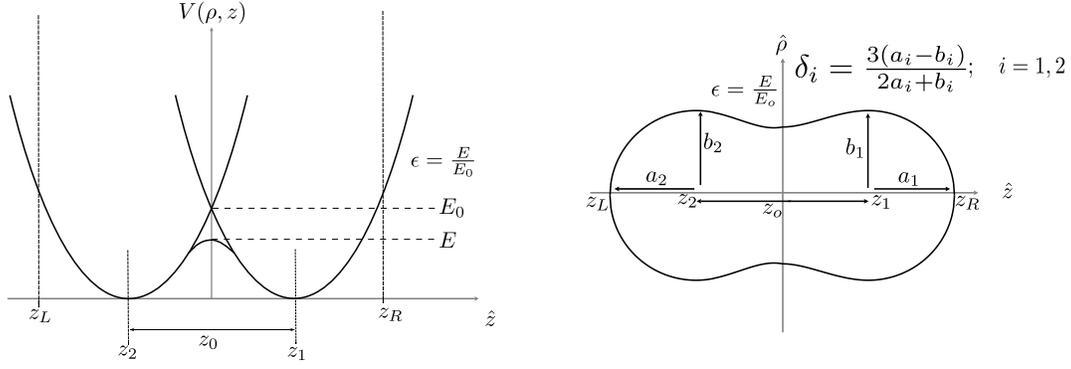}
\caption{\label{potential} Left: Two centre shell model (TCSM) potential \cite{MaruhnGreiner1972}. Right: Shape parameterizations based off the two centre shell model.}
\end{figure}

The 4-D Langevin approach describes the time evolution of the shape of nuclear surface defined by the TCSM collective variables, $q_{\mu} =( z_0/R_0, \delta_1, \delta_2, \alpha)$. 
These collective variables are depicted in Fig.~\ref{potential}. There $z_0/R_0$ refers to the distance between the potential minimum of the left and right fragments,
%which are then normalized with 
where $R_0=1.2\sqrt[3]{A}$ is the radius of spherical nucleus with mass number $A$. The parameters $\delta_i= 3(a_i-b_i)/(2a_i+b_i)$,  where i=\{1,2\}, describe the deformation of the right and left fragment tips. 
%In 3-D Langevin calculations \cite{usang2016,usang2017,aritomo2013,aritomo2014}, $\delta_1=\delta_2$ are implemented, but by increasing the degree of freedom by allowing $\delta_1$ and $\delta_2$ to evolve has allowed us to observe the emergence of some phenomena usually seen in experiments. 
The fourth parameter $q_4$ is the mass asymmetry $\alpha=(A_1-A_2)/(A_1+A_2)$ and it depends on the volumes of system to the left and right from the point $z=0$. The fifth parameter of TCSM shape parameterization $\epsilon$, defined as the ratio of actual and oscillator potentials at $z=0, \epsilon\equiv E/E_0$, see Fig. \ref{potential}, was kept constant, $\epsilon=0.35$, in all our calculations.

The time dependence of collective variables $q_{\mu}$ and the conjugated momenta $p_{mu}$ is described by the system of first order differential equations (Langevin equations),
\begin{align}\label{langevin}
\frac{dq_\mu}{dt}=&\left(m^{-1} \right)_{\mu \nu} p_\nu , \\
\frac{dp_\mu}{dt}=&-\frac{\partial U(q)}{\partial q_\mu} - \frac{1}{2}\frac{\partial m^{-1} _{\nu \sigma} }{\partial q_\mu} p_\nu p_\sigma
 -\gamma_{\mu \nu} m^{-1}_{\nu \sigma} p_\sigma +g_{\mu \nu} R_\nu (t), \nonumber
\end{align}
where the sums over the repeated indices are assumed. In Eqs. (\ref{langevin}) $U(q)$ is the potential energy, the 
$\gamma_{\mu\nu}$  and $(m)^{-1}_{\mu\nu}$ are the friction and inverse mass tensors and $g_{\mu\nu}R_\nu$ is the random force.

The potential energy $U(q)$ is calculated as the sum of liquid drop deformation energy \cite{krap79} and shell corrections \cite{strut67,strut68}. The single particle energies are calculated with the deformed Woods-Saxon potential \cite{pashkevich2008} fitted to the aforementioned TCSM shape parameterizations.% as was outlined in our previous paper \cite{lang4D}. 
The collective inertia tensor $m_{\mu\nu}$ is calculated based on the Werner-Wheeler approximation of the liquid drop mass tensor \cite{massliquid}.
The friction tensor $\gamma_{\mu\nu}$ is calculated from the wall-window friction formulation \cite{wallfriction,adeev,krapom,swiat1984,nixsierk}.
The random force are calculated as the product white noise $R_\nu$ and random force strength $g_{\mu\nu}$. More details are specified in our previous publications \cite{lang4D, usang-theory4}. 

A single event of Langevin calculation typically begins from the second minimum or in the vicinity of it.
%\textcolor{blue}{is it true? what you do if there is no second minimum? In the results that I got from you for Fm-256 and Fm-258 the calculations start from the saddle} of the potential energy surface until we reach the configuration when $R_\mathrm{neck}=0$ (scission configuration). 
If it fails to achieve scission configuration after 10,000 fm/c, the calculation is terminated. Typically 500,000 events are done per nuclei but we occasionally increase the number of events if it was a critical calculation. 
%Each successful fission events are counted and then tabulated with respect to the TKE and $\{q_\mu\}$, allowing us to create the TKE profiles for  the fission fragments of each nuclei. 

\section*{Summary}

%The Discussion should be succinct and must not contain subheadings.
By describing the fission process in terms of 4-D Langevin equations we have shown that the anomalously high TKE seen in $^{258}$Fm, $^{259}$Fm and $^{260}$Md is the inevitable results of splitting of nucleus into two almost double magic fragments that are symmetric as have been speculated by Hoffman \textit{et al.}\cite{hoffman1980}. However, unlike $^{236}$U that demonstrate super-long fission modes for the symmetric splitting, these three nuclei of $^{258}$Fm, $^{259}$Fm, $^{260}$Md and other fissioning systems around them have super-short fission modes.
The differences between super-long fission modes and the super-short fission modes shows itself in the Coulomb repulsion energy between the fission fragments that is stronger in the latter fission mode due to its shorter shape. 
Then a further investigation revealed that the super-short fission modes are present for all nuclei heavier than Einsteinium. Hence, the mystery on why the super-short fission modes are preferred in $^{258}$Fm, $^{259}$Fm, $^{260}$Md are solved.
We even see the slow disappearance of super-long fission modes, prominent in $^{236}$U and hardly identifiable for $^{250}$Cf. 
The present limitation of the calculation is due to the use of macroscopic transport coefficients that causes a slight underprediction of $\left<\mathrm{TKE}\right>$ for $^{258}$Fm, $^{259}$Fm, $^{260}$Md. Nevertheless, our analysis of $\delta$ distribution indicates that the allowance for $\delta$ in our 4-D Langevin calculation showed us multiple fission paths dependent on the combinations of $\delta_L$ and $\delta_H$. This has allowed us to explain how these transitions happen in a consistent manner.

\section*{Data availability}
All data generated or analyzed during this study are included in this published article (and its Supplementary Information files).

\bibliography{stdref}

\newcommand{\noop}[1]{}
\begin{thebibliography}{10}
\urlstyle{rm}
\expandafter\ifx\csname url\endcsname\relax
  \def\url#1{\texttt{#1}}\fi
\expandafter\ifx\csname urlprefix\endcsname\relax\def\urlprefix{URL }\fi
\expandafter\ifx\csname doiprefix\endcsname\relax\def\doiprefix{DOI: }\fi
\providecommand{\bibinfo}[2]{#2}
\providecommand{\eprint}[2][]{\url{#2}}

\bibitem{WangYe2018}
\bibinfo{author}{Wang, N.} \& \bibinfo{author}{Ye, W.}
\newblock \bibinfo{journal}{\bibinfo{title}{Probing nuclear dissipation with
  first-chance fission probability}}.
\newblock {\emph{\JournalTitle{Phys. Rev. C}}} \textbf{\bibinfo{volume}{97}},
  \bibinfo{pages}{014603}, \doiprefix\url{10.1103/PhysRevC.97.014603}
  (\bibinfo{year}{2018}).

\bibitem{Eslamizadeh2018}
\bibinfo{author}{Eslamizadeh, H.} \& \bibinfo{author}{Abdollahi, N.}
\newblock \bibinfo{journal}{\bibinfo{title}{Influence of the asymmetry
  parameter and dissipation coefficient of the $k$ coordinate on different
  aspects of fission of excited compound nuclei}}.
\newblock {\emph{\JournalTitle{Phys. Rev. C}}} \textbf{\bibinfo{volume}{97}},
  \bibinfo{pages}{024614}, \doiprefix\url{10.1103/PhysRevC.97.024614}
  (\bibinfo{year}{2018}).

\bibitem{usang2017}
\bibinfo{author}{Usang, M.~D.}, \bibinfo{author}{Ivanyuk, F.~A.},
  \bibinfo{author}{Ishizuka, C.} \& \bibinfo{author}{Chiba, S.}
\newblock \bibinfo{journal}{\bibinfo{title}{Analysis of the total kinetic
  energy of fission fragments with the langevin equation}}.
\newblock {\emph{\JournalTitle{Phys. Rev. C}}} \textbf{\bibinfo{volume}{96}},
  \bibinfo{pages}{064617}, \doiprefix\url{10.1103/PhysRevC.96.064617}
  (\bibinfo{year}{2017}).

\bibitem{Sierk2017}
\bibinfo{author}{Sierk, A.~J.}
\newblock \bibinfo{journal}{\bibinfo{title}{Langevin model of low-energy
  fission}}.
\newblock {\emph{\JournalTitle{Phys. Rev. C}}} \textbf{\bibinfo{volume}{96}},
  \bibinfo{pages}{034603}, \doiprefix\url{10.1103/PhysRevC.96.034603}
  (\bibinfo{year}{2017}).

\bibitem{mazurek2017}
\bibinfo{author}{Mazurek, K.}, \bibinfo{author}{Nadtochy, P.~N.},
  \bibinfo{author}{Ryabov, E.~G.} \& \bibinfo{author}{Adeev, G.~D.}
\newblock \bibinfo{journal}{\bibinfo{title}{Fission fragment distributions
  within dynamical approach}}.
\newblock {\emph{\JournalTitle{Eur. Phys. J. A}}}
  \textbf{\bibinfo{volume}{53}}, \bibinfo{pages}{79},
  \doiprefix\url{10.1140/epja/i2017-12262-1} (\bibinfo{year}{2017}).

\bibitem{eslamizadeh2017}
\bibinfo{author}{Eslamizadeh, H.}
\newblock \bibinfo{journal}{\bibinfo{title}{Theoretical study of different
  features of the fission process of excited nuclei in the framework of the
  modified statistical model and four-dimensional dynamical model}}.
\newblock {\emph{\JournalTitle{J. Phys. G: Nucl. Part. Phys.}}}
  \textbf{\bibinfo{volume}{44}}, \bibinfo{pages}{025102}
  (\bibinfo{year}{2017}).

\bibitem{usang2016}
\bibinfo{author}{Usang, M.~D.}, \bibinfo{author}{Ivanyuk, F.~A.},
  \bibinfo{author}{Ishizuka, C.} \& \bibinfo{author}{Chiba, S.}
\newblock \bibinfo{journal}{\bibinfo{title}{Effects of microscopic transport
  coefficients on fission observables calculated by the langevin equation}}.
\newblock {\emph{\JournalTitle{Phys. Rev. C}}} \textbf{\bibinfo{volume}{94}},
  \bibinfo{pages}{044602}, \doiprefix\url{10.1103/PhysRevC.94.044602}
  (\bibinfo{year}{2016}).

\bibitem{pahlavani2015}
\bibinfo{author}{Pahlavani, M.~R.} \& \bibinfo{author}{Mirfathi, S.~M.}
\newblock \bibinfo{journal}{\bibinfo{title}{Dynamics of neutron-induced fission
  of $^{235}\mathrm{U}$ using four-dimensional langevin equations}}.
\newblock {\emph{\JournalTitle{Phys. Rev. C}}} \textbf{\bibinfo{volume}{92}},
  \bibinfo{pages}{024622}, \doiprefix\url{10.1103/PhysRevC.92.024622}
  (\bibinfo{year}{2015}).

\bibitem{asano2004}
\bibinfo{author}{Asano, T.} \emph{et~al.}
\newblock \bibinfo{journal}{\bibinfo{title}{{Dynamical calculation of
  multi-modal nuclear fission of fermium nuclei}}}.
\newblock {\emph{\JournalTitle{J. Nucl. Radiochem. Sci.}}}
  \textbf{\bibinfo{volume}{5}}, \bibinfo{pages}{1--5} (\bibinfo{year}{2004}).

\bibitem{karpov2001}
\bibinfo{author}{Karpov, A.~V.}, \bibinfo{author}{Nadtochy, P.~N.},
  \bibinfo{author}{Vanin, D.~V.} \& \bibinfo{author}{Adeev, G.~D.}
\newblock \bibinfo{journal}{\bibinfo{title}{Three-dimensional langevin
  calculations of fission fragment mass-energy distribution from excited
  compound nuclei}}.
\newblock {\emph{\JournalTitle{Phys. Rev. C}}} \textbf{\bibinfo{volume}{63}},
  \bibinfo{pages}{054610}, \doiprefix\url{10.1103/PhysRevC.63.054610}
  (\bibinfo{year}{2001}).

\bibitem{wada93}
\bibinfo{author}{Wada, T.}, \bibinfo{author}{Abe, Y.} \&
  \bibinfo{author}{Carjan, N.}
\newblock \bibinfo{journal}{\bibinfo{title}{One-body dissipation in agreement
  with prescission neutrons and fragment kinetic energies}}.
\newblock {\emph{\JournalTitle{Phys. Rev. Lett.}}}
  \textbf{\bibinfo{volume}{70}}, \bibinfo{pages}{3538--3541},
  \doiprefix\url{10.1103/PhysRevLett.70.3538} (\bibinfo{year}{1993}).

\bibitem{aritomo2013}
\bibinfo{author}{Aritomo, Y.} \& \bibinfo{author}{Chiba, S.}
\newblock \bibinfo{journal}{\bibinfo{title}{{Fission process of nuclei at low
  excitation energies with a Langevin approach}}}.
\newblock {\emph{\JournalTitle{Phys. Rev. C}}} \textbf{\bibinfo{volume}{88}},
  \bibinfo{pages}{044614} (\bibinfo{year}{2013}).

\bibitem{aritomo2014}
\bibinfo{author}{Aritomo, Y.}, \bibinfo{author}{Chiba, S.} \&
  \bibinfo{author}{Ivanyuk, F.}
\newblock \bibinfo{journal}{\bibinfo{title}{{Fission dynamics at low excitation
  energy}}}.
\newblock {\emph{\JournalTitle{Phys. Rev. C}}} \textbf{\bibinfo{volume}{90}},
  \bibinfo{pages}{054609} (\bibinfo{year}{2014}).

\bibitem{hoffman1980}
\bibinfo{author}{Hoffman, D.~C.} \emph{et~al.}
\newblock \bibinfo{journal}{\bibinfo{title}{12.3-min $^{256}\mathrm{Cf}$ and
  43-min $^{258}\mathrm{Md}$ and systematics of the spontaneous fission
  properties of heavy nuclides}}.
\newblock {\emph{\JournalTitle{Phys. Rev. C}}} \textbf{\bibinfo{volume}{21}},
  \bibinfo{pages}{972--981}, \doiprefix\url{10.1103/PhysRevC.21.972}
  (\bibinfo{year}{1980}).

\bibitem{hulet1989}
\bibinfo{author}{Hulet, E.~K.} \emph{et~al.}
\newblock \bibinfo{journal}{\bibinfo{title}{Spontaneous fission properties of
  $^{258}\mathrm{Fm}$, $^{259}\mathrm{Md}$, $^{260}\mathrm{Md}$,
  $^{258}\mathrm{No}$, and $^{260}]$: Bimodal fission}}.
\newblock {\emph{\JournalTitle{Phys. Rev. C}}} \textbf{\bibinfo{volume}{40}},
  \bibinfo{pages}{770--784}, \doiprefix\url{10.1103/PhysRevC.40.770}
  (\bibinfo{year}{1989}).

\bibitem{hoffman1990-sf}
\bibinfo{author}{Hoffman, D.~C.} \emph{et~al.}
\newblock \bibinfo{journal}{\bibinfo{title}{Spontaneous fission properties of
  2.9-s $^{256}\mathrm{No}$}}.
\newblock {\emph{\JournalTitle{Phys. Rev. C}}} \textbf{\bibinfo{volume}{41}},
  \bibinfo{pages}{631--639}, \doiprefix\url{10.1103/PhysRevC.41.631}
  (\bibinfo{year}{1990}).

\bibitem{brosa1990}
\bibinfo{author}{Brosa, U.}, \bibinfo{author}{Grossmann, S.} \&
  \bibinfo{author}{M{\"u}ller, A.}
\newblock \bibinfo{journal}{\bibinfo{title}{Nuclear scission}}.
\newblock {\emph{\JournalTitle{Phys. Rep.}}} \textbf{\bibinfo{volume}{197}},
  \bibinfo{pages}{167 -- 262},
  \doiprefix\url{http://dx.doi.org/10.1016/0370-1573(90)90114-H}
  (\bibinfo{year}{1990}).

\bibitem{lang4D}
\bibinfo{author}{Ishizuka, C.} \emph{et~al.}
\newblock \bibinfo{journal}{\bibinfo{title}{Four-dimensional langevin approach
  to low-energy nuclear fission of $^{236}\mathbf{U}$}}.
\newblock {\emph{\JournalTitle{Phys. Rev. C}}} \textbf{\bibinfo{volume}{96}},
  \bibinfo{pages}{064616}, \doiprefix\url{10.1103/PhysRevC.96.064616}
  (\bibinfo{year}{2017}).

\bibitem{usang-theory4}
\bibinfo{author}{Usang, M.~D.}, \bibinfo{author}{Ivanyuk, F.~A.},
  \bibinfo{author}{Ishizuka, C.}, \bibinfo{author}{Chiba, S.} \&
  \bibinfo{author}{Maruhn, J.~A.}
\newblock \bibinfo{journal}{\bibinfo{title}{Fission observables from {4D}
  langevin calculations with macroscopic transport coefficients}}.
\newblock {\emph{\JournalTitle{EPJ Web Conf.}}} \textbf{\bibinfo{volume}{169}},
  \bibinfo{pages}{00027}, \doiprefix\url{10.1051/epjconf/201816900027}
  (\bibinfo{year}{2018}).

\bibitem{viola1985}
\bibinfo{author}{Viola, V.~E.}, \bibinfo{author}{Kwiatkowski, K.} \&
  \bibinfo{author}{Walker, M.}
\newblock \bibinfo{journal}{\bibinfo{title}{{Systematics of fission fragment
  total kinetic energy release}}}.
\newblock {\emph{\JournalTitle{Phys. Rev. C}}} \textbf{\bibinfo{volume}{31}},
  \bibinfo{pages}{1550} (\bibinfo{year}{1985}).

\bibitem{unik1971}
\bibinfo{author}{Unik, J.~P.} \& \bibinfo{author}{Gindler, J.~E.}
\newblock \bibinfo{title}{Critical review of the energy released in nuclear
  fission anl-7748.}
\newblock \bibinfo{type}{Tech. Rep.}, \bibinfo{institution}{Argonne National
  Lab., Ill.} (\bibinfo{year}{1971}).

\bibitem{hoffman1995}
\bibinfo{author}{Hoffman, D.~C.} \& \bibinfo{author}{Lane, M.~R.}
\newblock \bibinfo{journal}{\bibinfo{title}{Spontaneous fission}}.
\newblock {\emph{\JournalTitle{Radiochim. Acta}}}
  \textbf{\bibinfo{volume}{70}}, \bibinfo{pages}{135--146}
  (\bibinfo{year}{1995}).

\bibitem{hoffman1996}
\bibinfo{author}{Hoffman, D.~C.}, \bibinfo{author}{Hamilton, T.~M.} \&
  \bibinfo{author}{Lane, M.~R.}
\newblock \emph{\bibinfo{title}{{Nuclear Decay Modes}}}
  (\bibinfo{publisher}{CRC}, \bibinfo{address}{Boca Raton, FL},
  \bibinfo{year}{1996}).

\bibitem{JENDL}
\bibinfo{author}{Shibata, K.} \emph{et~al.}
\newblock \bibinfo{journal}{\bibinfo{title}{{JENDL-4.0: A New Library for
  Nuclear Science and Engineering}}}.
\newblock {\emph{\JournalTitle{J. Nucl. Sci. Technol.}}}
  \textbf{\bibinfo{volume}{48}}, \bibinfo{pages}{1--30},
  \doiprefix\url{10.1080/18811248.2011.9711675} (\bibinfo{year}{2011}).

\bibitem{moody}
\bibinfo{author}{Moody, K.} \emph{et~al.}
\newblock \bibinfo{journal}{\bibinfo{title}{Decay properties of heavy
  mendelevium isotopes}}.
\newblock {\emph{\JournalTitle{Nuclear Physics A}}}
  \textbf{\bibinfo{volume}{563}}, \bibinfo{pages}{21 -- 73},
  \doiprefix\url{https://doi.org/10.1016/0375-9474(93)90010-U}
  (\bibinfo{year}{1993}).

\bibitem{MaruhnGreiner1972}
\bibinfo{author}{Maruhn, J.} \& \bibinfo{author}{Greiner, W.}
\newblock \bibinfo{journal}{\bibinfo{title}{{The asymmetrie two center shell
  model}}}.
\newblock {\emph{\JournalTitle{Z. Phys.}}} \textbf{\bibinfo{volume}{251}},
  \bibinfo{pages}{431--457} (\bibinfo{year}{1972}).

\bibitem{krap79}
\bibinfo{author}{Krappe, H.~J.}, \bibinfo{author}{Nix, J.~R.} \&
  \bibinfo{author}{Sierk, A.~J.}
\newblock \bibinfo{journal}{\bibinfo{title}{{Unified nuclear potential for
  heavy-ion elastic scattering, fusion, fission, and ground-state masses and
  deformations}}}.
\newblock {\emph{\JournalTitle{Phys. Rev. C}}} \textbf{\bibinfo{volume}{20}},
  \bibinfo{pages}{992--1013}, \doiprefix\url{10.1103/PhysRevC.20.992}
  (\bibinfo{year}{1979}).

\bibitem{strut67}
\bibinfo{author}{Strutinsky, V.}
\newblock \bibinfo{journal}{\bibinfo{title}{{Shell effects in nuclear masses
  and deformation energies}}}.
\newblock {\emph{\JournalTitle{Nucl. Phys. A}}} \textbf{\bibinfo{volume}{95}},
  \bibinfo{pages}{420--442}, \doiprefix\url{10.1016/0375-9474(67)90510-6}
  (\bibinfo{year}{1967}).

\bibitem{strut68}
\bibinfo{author}{Strutinsky, V.}
\newblock \bibinfo{journal}{\bibinfo{title}{{``Shells'' in deformed nuclei}}}.
\newblock {\emph{\JournalTitle{Nucl. Phys. A}}} \textbf{\bibinfo{volume}{122}},
  \bibinfo{pages}{1--33}, \doiprefix\url{10.1016/0375-9474(68)90699-4}
  (\bibinfo{year}{1968}).

\bibitem{pashkevich2008}
\bibinfo{author}{Pashkevich, V.} \& \bibinfo{author}{Rusanov, A.}
\newblock \bibinfo{journal}{\bibinfo{title}{The 226th fission valleys}}.
\newblock {\emph{\JournalTitle{Nucl. Phys. A}}} \textbf{\bibinfo{volume}{810}},
  \bibinfo{pages}{77 -- 90},
  \doiprefix\url{http://dx.doi.org/10.1016/j.nuclphysa.2008.06.013}
  (\bibinfo{year}{2008}).

\bibitem{massliquid}
\bibinfo{author}{Davies, K. T.~R.}, \bibinfo{author}{Sierk, A.~J.} \&
  \bibinfo{author}{Nix, J.~R.}
\newblock \bibinfo{journal}{\bibinfo{title}{{Effect of viscosity on the
  dynamics of fission}}}.
\newblock {\emph{\JournalTitle{Phys. Rev. C}}} \textbf{\bibinfo{volume}{13}},
  \bibinfo{pages}{2385--2403}, \doiprefix\url{10.1103/PhysRevC.13.2385}
  (\bibinfo{year}{1976}).

\bibitem{wallfriction}
\bibinfo{author}{Blocki, J.} \emph{et~al.}
\newblock \bibinfo{journal}{\bibinfo{title}{{One-body dissipation and the
  super-viscidity of nuclei}}}.
\newblock {\emph{\JournalTitle{Ann. Phys.}}} \textbf{\bibinfo{volume}{113}},
  \bibinfo{pages}{330--386} (\bibinfo{year}{1978}).

\bibitem{adeev}
\bibinfo{author}{Adeev, G.}, \bibinfo{author}{Karpov, A.},
  \bibinfo{author}{Nadtochii, P.} \& \bibinfo{author}{Vanin, D.}
\newblock \bibinfo{journal}{\bibinfo{title}{{Multidimensional Stochastic
  Approach to the Fission Dynamics of Excited Nuclei}}}.
\newblock {\emph{\JournalTitle{Phys. Part. Nucl.}}}
  \textbf{\bibinfo{volume}{36}}, \bibinfo{pages}{378--426}
  (\bibinfo{year}{2005}).

\bibitem{krapom}
\bibinfo{author}{Krappe, H.~J.} \& \bibinfo{author}{Pomorski, K.}
\newblock \emph{\bibinfo{title}{{Theory of Nuclear Fission: A Textbook}}}, vol.
  \bibinfo{volume}{838} of \emph{\bibinfo{series}{Lecture Notes in Physics}}
  (\bibinfo{publisher}{Springer}, \bibinfo{address}{New York},
  \bibinfo{year}{2012}).

\bibitem{swiat1984}
\bibinfo{author}{Swiatecki, W.}
\newblock \bibinfo{journal}{\bibinfo{title}{{Macroscopic treatment of nuclear
  dynamics}}}.
\newblock {\emph{\JournalTitle{Nucl. Phys. A}}} \textbf{\bibinfo{volume}{428}},
  \bibinfo{pages}{199--222}, \doiprefix\url{10.1016/0375-9474(84)90252-5}
  (\bibinfo{year}{1984}).

\bibitem{nixsierk}
\bibinfo{author}{Nix, J.~R.} \& \bibinfo{author}{Sierk, A.~J.}
\newblock \bibinfo{journal}{\bibinfo{title}{{Dynamics of fission and heavy ion
  reactions}}}.
\newblock {\emph{\JournalTitle{Nucl. Phys. A}}} \textbf{\bibinfo{volume}{428}},
  \bibinfo{pages}{161--175}, \doiprefix\url{10.1016/0375-9474(84)90249-5}
  (\bibinfo{year}{1984}).

\bibitem{thesis-mark}
\bibinfo{author}{Usang, M.~D.}
\newblock \emph{\bibinfo{title}{Study on mechanisms of nuclear fission by
  Langevin equation}}.
\newblock Ph.D. thesis, \bibinfo{school}{Tokyo Institute of Technology},
  \bibinfo{address}{Tokyo, Japan} (\bibinfo{year}{2018}).

\end{thebibliography}

%\noindent LaTeX formats citations and references automatically using the bibliography records in your .bib file, which you can edit via the project menu. Use the cite command for an inline citation, e.g.  \cite{Hao:gidmaps:2014}.

%For data citations of datasets uploaded to e.g. \emph{figshare}, please use the \verb|howpublished| option in the bib entry to specify the platform and the link, as in the \verb|Hao:gidmaps:2014| example in the sample bibliography file.

\section*{Acknowledgements}

This study comprises the results of “Research and development of an innovative transmutation
system of LLFP by fast reactors” (18K03642) entrusted to the Tokyo Institute of Technology by the Ministry of Education,
Culture, Sports, Science and Technology of Japan (MEXT).
We would also like to acknowledge the IAEA CRP on beta-delayed neutrons (F41030).

\section*{Author contributions statement}

%Must include all authors, identified by initials, for example:
S.C. conceived the work(s), S.C. provided the Langevin and analysis code, F.A.I. provided the code for calculating potential energy surface and macroscopic transport coefficients, M.D.U. calculated the results(s) and revised the analysis code, M.D.U. and C.I. analyzed the results.  All authors reviewed the manuscript. 

\section*{Additional information}

The author(s) declare no competing interests. 
Most of the figures had been included for M.D.U. PhD thesis manuscript \cite{thesis-mark} prior to the submission of this paper.

%To include, in this order: \textbf{Accession codes} (where applicable); \textbf{Competing interests} (mandatory statement). 

%The corresponding author is responsible for submitting a \href{http://www.nature.com/srep/policies/index.html#competing}{competing interests statement} on behalf of all authors of the paper. This statement must be included in the submitted article file.

% \begin{figure}[ht]
% \centering
% \includegraphics[width=\linewidth]{stream}
% \caption{Legend (350 words max). Example legend text.}
% \label{fig:stream}
% \end{figure}

% \begin{table}[ht]
% \centering
% \begin{tabular}{|l|l|l|}
% \hline
% Condition & n & p \\
% \hline
% A & 5 & 0.1 \\
% \hline
% B & 10 & 0.01 \\
% \hline
% \end{tabular}
% \caption{\label{tab:example}Legend (350 words max). Example legend text.}
% \end{table}

% Figures and tables can be referenced in LaTeX using the ref command, e.g. Figure \ref{fig:stream} and Table \ref{tab:example}.

\end{document}